\documentclass[useAMS,usenatbib]{mn2e}
\usepackage{amsmath}
\usepackage{comment}
\usepackage{graphicx}
\usepackage{rotating}
\usepackage{color}
\usepackage{aas_macros}

\newcommand{\beq}{\begin{equation}}
\newcommand{\enq}{\end{equation}}
\newcommand{\beqa}{\begin{eqnarray}}
\newcommand{\enqa}{\end{eqnarray}}
\newcommand{\beit}{\begin{itemize}}
\newcommand{\enit}{\end{itemize}}
\newcommand{\bem}{\begin{pmatrix}}
\newcommand{\enm}{\end{pmatrix}}

\newcommand{\lat}{\left\langle}
\newcommand{\rat}{\right\rangle}
\newcommand{\av}[1]{\lat #1 \rat}

\newcommand{\lb}{\left [}
\newcommand{\rb}{\right ]}
\newcommand{\lp}{\left (}
\newcommand{\rp}{\right )}

\newcommand{\bes}{\begin{sideways}}
\newcommand{\ees}{\end{sideways}}

\newcommand{\Nca}{\bar N \sigma^2_A}

\newcommand{\mpc}{ {h^{-1}\rm Mpc }}

\title[Observables for large scale structure]{Sufficient observables for large scale structure in galaxy surveys}
\author[Carron and Szapudi]{J. Carron\thanks{E-mail:
carron@ifa.hawaii.edu} and I. Szapudi  \\
Institute for Astronomy, University of Hawaii, 2680 Woodlawn Drive, Honolulu, HI, 96822}
\begin{document}

\date{\today}

\pagerange{\pageref{firstpage}--\pageref{lastpage}} \pubyear{2013}

\maketitle

\label{firstpage}

\begin{abstract} 
Beyond the linear regime, the power spectrum and higher order moments of the matter field no longer capture all cosmological information encoded in density fluctuations. While non-linear transforms have been proposed to extract this information lost to traditional methods, up to now, the way to generalize these techniques to discrete processes was unclear; ad hoc extensions had some success. We pointed out in \cite{CarronSzapudi13} that the logarithmic transform approximates extremely well the optimal ``sufficient statistics'', observables that extract all information from the (continuous) matter field. Building on these results, we generalize optimal transforms to discrete galaxy fields. We focus our calculations on the Poisson sampling of an underlying lognormal density field. 
We solve and test the one-point case in detail, and sketch out the sufficient observables for the multi-point case. Moreover, we present an accurate approximation to the sufficient observables in terms of the mean and spectrum of a non-linearly transformed field.
We find that the corresponding optimal non-linear transformation is directly related to the maximum a posteriori Bayesian reconstruction of the underlying continuous field with a lognormal prior as put forward in \cite{KitauraEtal10}. Thus simple recipes for realizing the sufficient observables can be built on previously proposed algorithms that have been successfully implemented and tested in simulations. 
\end{abstract}

\begin{keywords}{large-scale structure of Universe, cosmology: theory, methods: statistical} 
\end{keywords}
\section{Introduction}
In the current inflationary paradigm, the small primordial density fluctuations are believed to be very close to a Gaussian field. The natural descriptors of such fields are two-point correlation functions, or (power) spectra in Fourier space. However, it was established first in $N$-body simulations \citep{RimesHamilton05,RimesHamilton06}, and subsequently with analytical calculations \citep{NeyrinckEtal06} that the spectrum of the matter field loses its effectiveness as the fluctuations grow. Fourier modes of the density become strongly coupled \citep{MeiksinWhite99,ScoccimaroEtal99}, resulting in large covariances effectively diminishing the available information. In this context, a variety of non-linear transformations of the field such as Gaussianization \citep{Weinberg92,NeyrinckEtal11,ZhangEtal11,YuEtal11}, the logarithmic mapping \citep{NeyrinckEtal09,SeoEtal11,SeoEtal12,Carron12}, Box-Cox transformations \citep{JoachimiEtal11} or clipping \citep{SimpsonEtal11,SimpsonEtal13}, have been shown to increase the fidelity to linear theory and/or recapture information in the noise free fields.  These results can be understood within the simple yet qualitatively and to some extent quantitatively accurate lognormal model \citep{1991MNRAS.248....1C,2001ApJ...561...22K} of the statistics of the matter field. 
It has been shown that the full set of $N$-point moments of fields of this type carry very little information in the high variance regime \citep[see][and references therein for an extensive discussion on these statistical issues]{Carron:2012qf}. 
\newline
\indent
Before these ideas can applied to extract more information from galaxy surveys, a key issue to be dealt with is discreteness. Galaxy fields correspond to a set of points rather than a continuous random field, and it is not entirely clear how and to what extent the methods and conclusions of these works apply. While studies such as \cite{NeyrinckEtal11} suggests that analogous methods should still bring some improvement, up to now the estimators relied on ad hoc generalizations of the logarithmic and similar transforms to discrete data sets. 
\newline
\indent
In \cite{CarronSzapudi13} we have shown rigorously that such non-linear transforms of the matter field typically work because they i) undo (some of) the non-linear evolution, thus ii) Gaussianize the distribution, and last but not least iii) they correspond to a good approximation to sufficient statistics, observables that extract all available information from the matter field on a given cosmological parameter. 
The logarithmic transformation would correspond to exact sufficient statistics if the underlying continuous field were lognormal. While the evolved dark matter field is only approximately lognormal, apparently it is close enough that the amount of information extracted by the simple logarithmic transformation is virtually indistinguishable from that of the exact (and vastly more complicated) sufficient statistics. Thus a simple way presents itself to generalize these findings to the discrete galaxy fields in surveys: assume a lognormal field sampled in a Poisson fashion, and construct the corresponding sufficient statistics. Both of these assumptions are approximate, 
 nevertheless, we expect these results to capture the essential properties of dark matter fields encoded by the distribution of galaxies. Also, it will be clear how our results generalize to more complex statistical models.
\section{Methods} \label{section1}
We build on our previous work \citep{CarronSzapudi13} to which we refer for details, but can be summarized as follows. Let $p(\delta)$ be the one-point PDF of the fluctuation $\delta$. It is well known that the statistic
\beq
o(\delta) = \delta^2
\enq 
(i.e. the variance) contains the entire Fisher information $\av{\lp \partial_\alpha \ln p \rp^2}$ of $p(\delta) $ whenever the probability density $p(\delta)$ is Gaussian. 
This is a special case of a more general identity, valid for  any $p(\delta)$: the observable
\beq \label{optimalgeneral}
o(\delta) = c_1 \frac{\partial \ln p(\delta)}{\partial \alpha} + c_2
\enq
carries the entire Fisher information content of $p(\delta)$ on the parameter $\alpha$ (in this equation $c_1$ and $c_2$ are arbitrary constants). These statistics can be therefore be considered 'sufficient', or optimal, with respect to their Fisher information content. They can be read directly from the shape of the PDF.
\newline
\indent
In this paper we construct the corresponding sufficient observables for the PDF of a discrete galaxy field, given schematically as
\beq\label{optimalgeneralPN}
o(N) =  c_1 \frac{\partial \ln P(N)}{\partial \alpha} + c_2,
\enq
where $P(N)$ is the probability of observing $N = 0,1,...$ galaxies in a given cell. Initially, we restrict our analysis to one-point probabilities and one-point optimal observables, to obtain local transformations exactly analogous to previous successful methods. Our focus on the one-point PDF makes the variance of the fluctuations the only model parameter of relevance. Nevertheless, generalizations to multipoint probabilities $P(N_1,N_2,\cdots)$ will be obvious after the detailed calculation. 
\newline
\indent
We will use the local Poisson model as a natural way for discrete sampling of an underlying continuous distribution. Less obvious is the choice of the underlying PDF $p(\delta)$. In  \citep{CarronSzapudi13} we used results from perturbation theory on the moments of the dark mater $\delta$ field to show that
\beq \label{optimalln}
o(\delta) = \ln^2(1 + \delta)
\enq
was a very good approximation to the optimal observable when the index $n$ of the power spectrum is reasonably close to $-1$. This result was very successfully tested against the Millennium Simulation \citep{SpringelEtal2005} density field. Since equation~\eqref{optimalln} is the optimal statistic~\eqref{optimalgeneral} for the lognormal PDF, this model presents itself as a plausible choice for the underlying continuous PDF for the construction of the sufficient observable.
\subsection{Local Poisson model with lognormal density}
Using the local Poisson model with lognormal underlying matter density, the probability $P(N)$ of observing $N = 0,1,...$ galaxies in a given cell is given by
\beq \label{Poisson}
P(N) = \int_{-1}^\infty d\delta \:p(\delta) p(N | \delta)
\enq
where
\beq
p(N | \delta) =\frac{1}{N!} e^{-\bar N \lp  1 + \delta \rp  }\lb \bar N \lp 1 + \delta \rp\rb^N,
\enq
and $\bar N = \av{N}$ is the mean number of galaxy in the cell. More realistic local models for discreteness including deviations from the Poissonity or biasing schemes can be implemented analogously \cite[e.g.]{KitauraEtal13}. The lognormal PDF of $\delta$ reads
\beq
p(\delta) = \frac{1}{\sqrt{2\pi \sigma^2_A} \lp1 + \delta \rp} \exp \lb -\frac 1{2\sigma^2_A}\lp \ln ( 1 + \delta )  - \bar A \rp^2 \rb.
\enq
Writing $A = \ln\lp 1 + \delta \rp$, then the parameters of the above equation are set by 
\beq
\begin{split}
\bar A &= \av{A} = -\frac 12 \sigma^2_A \\
\sigma^2_A &= \av{\lp A -\bar A \rp^2} = \ln \lp 1 + \sigma^2 \rp, 
\end{split}
\enq
where $\sigma^2 = \av{\delta^2}$.
The first relation ensures that $\delta$ has zero mean. The log-density $A$ has a Gaussian PDF.
\subsubsection*{Saddle-point approximation}
We found that $P(N)$ can be obtained with great accuracy via saddle-point approximation. Specifically,  we first write Eq. \eqref{Poisson} using $A = \ln \lp 1 + \delta \rp$ as
\beq \label{PN}
P(N) = \int_{-\infty}^\infty dA\:\exp \lb g_N(A) \rb,
\enq
with $g_N(A) = \ln p(A) + \ln P(N | A)$. 
We then approximate the integrand as
\beq
g_N(A) \simeq g_N(A^*) - \frac 12 (A - A^*)^2 |g_N''(A^*)| 
\enq
where $A^* = A^*(N)$ is the point where $g_N(A)$ is maximal. Eq. \eqref{PN} becomes a Gaussian integral resulting in
\beq \label{SP}
 P(N) \simeq \sqrt{\frac{2\pi}{|g_N''(A^*)| }} \exp \lp {g_N(A^*)} \rp.
\enq
Collecting the terms from the lognormal and Poisson PDFs, we have in our case
\beq
\begin{split}
g_N(A) &= -\frac 1{2\sigma^2_A} \lp A - \bar A\rp^2- \frac 12 \ln \lp 2\pi \sigma^2_A\rp    \\
&\quad-\bar N e^A + N\: A +  N\:\ln \bar N  - \ln N!
\end{split}
\enq
The saddle-point $A^*$ is obtained by setting $g_N'(A^*) = 0$, giving the following non-linear equation
\beq \label{saddlepoint}
e^{A^*} + \frac{A^*}{\Nca} = \frac{N-1/2}{\bar N}.
\enq
Equivalently
\beq
\delta^* =\delta_g - \frac{\ln \lp 1 + \delta^* \rp}{\Nca} -\frac{1}{2\bar N},
\enq
where $\delta_g = N/\bar N -1$.
The curvature term $g_N''(A^*)$ is
\beq
g_N''(A^*) = -\frac{1}{\sigma^2_A} -\bar N e^{A^*} = -\frac{1}{\sigma^2_A}\lb1   +  \Nca\lp 1 + \delta^* \rp \rb.
\enq
Equation \eqref{saddlepoint} is always well behaved with a unique solution. 

\subsection{The sufficient observable}
With the above results we proceed to construct the sufficient observable. Let us start with some general observations.  To get the optimal observable we need $\partial_\alpha \ln P(N)$. 
From Eq. \eqref{SP},
\beq \label{lnPN}
\ln P(N) = g_N(A^*) -\frac 12 \ln g''(A^*) + \frac 12 \ln \lp 2 \pi \rp.
\enq
When performing the derivatives, some care must be taken, as $\alpha$ enters in two different ways: in the density PDF $p(A)$ and in the solution to the saddle-point equation $A^*$. The first term in \eqref{lnPN} can be dealt with simply. We note that in general $g_N(A) = \ln p(A) + \ln P(N | A)$.  Thus, we can write
\beq \label{18}
\begin{split}
\frac{d g_N (A^*)}{d \alpha} &= \frac{\partial g_N(A^*)}{\partial A} \frac{\partial A^*}{\partial \alpha} + \frac{\partial g_N(A^*)}{\partial \alpha} \\
&= \frac{\partial \ln p(A^*)}{\partial \alpha} 
\end{split}
\enq
The first term in the upper equation vanishes by definition of the saddle point. The second term reduces to the right hand side of the lower equation since  $\ln P(N | A)$ carries no dependence on $\alpha$. Thus \eqref{18} is simply the optimal statistics of the underlying continuous density field evaluated at the point $A^*(N)$. To obtain the optimal statistics of the galaxy field we only need to add a correction from the curvature term. 
To write explicitly the curvature term, we need $\partial_\alpha A^*$. This is obtained taking the derivative of the saddle point equation $g'_N(A^*) = 0$, with the result
\beq
\lp \frac{\partial A^*}{\partial \alpha} \rp g_N''(A^*) = -\frac{\partial \ln p'(A^*)}{\partial \alpha}
\enq
We have all the ingredients to write down the optimal observable $o(N)$ for our Poisson lognormal model. Collecting the relevant terms  
we get after some algebra
\beq \label{Om}
\begin{split}
o(N) &= \ln^2(1 + \delta^*)  \\
&\quad - \frac{\ln \lp 1 + \sigma^2\rp \Nca(1 + \delta^*)}{1 +\Nca(1 + \delta^*)} \lp  1 + \frac{\ln(1 + \delta^*)}{1+ \Nca(1+ \delta^*)} \rp.
\end{split}
\enq
We present the interpretation of $\delta^*$ and of the second term in the above observable later on.
\section{Tests and results} \label{results}
To illustrate the behavior of $P(N)$ and test our statistics in the parameter space spanned by $\sigma^2$ and $\bar N$, we proceed as follows. Letting $R$ be the radius in $\mpc$ of a (spherical) cell, we set the expected power law behaviors
\beq \label{scaling}
\sigma^2 =  \sigma^2_8 \lp \frac{ R} {8\: \mpc}\rp^{-(n+3)},\quad \bar N = \bar N_8 \lp \frac{ R} {8\: \mpc}\rp^3,
\enq
where we use standard values $n = -1$, and $\sigma_8 = 0.8$. We will use for the purposes of this paper two different values of $\bar N_8$. The approximate SDSS LRGs density \citep{PercivalEtal07} $10^{-4} \lp h/\rm{Mpc} \rp^{-3}$  giving  $\bar N_8 = 0.2$, and a larger sampling rate $\bar N_8 = 1$ for comparison.
\subsubsection*{Saddle-point approximation }
\begin{figure}
\begin{center}
\includegraphics[width = 0.5\textwidth]{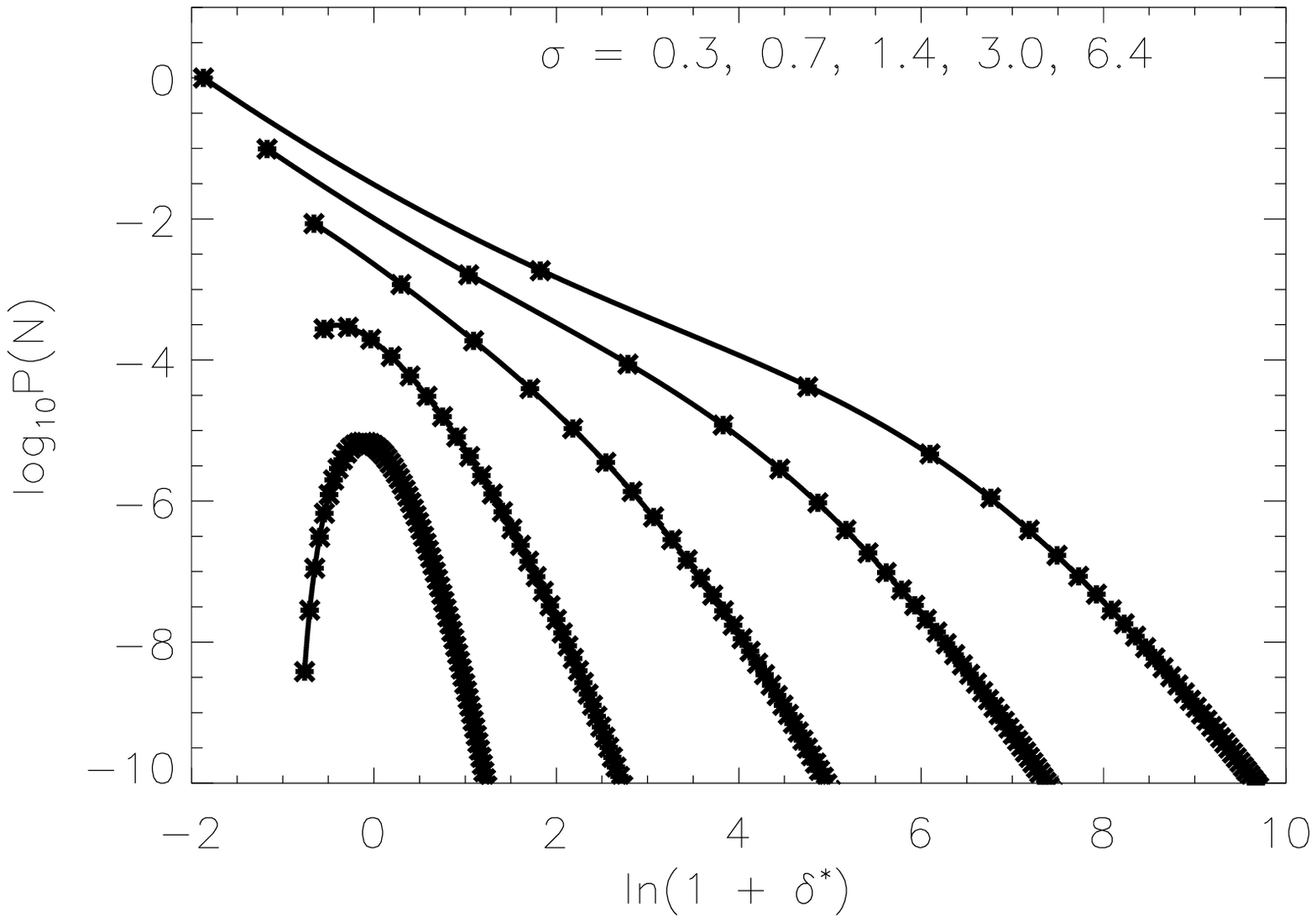}
\includegraphics[width = 0.5\textwidth]{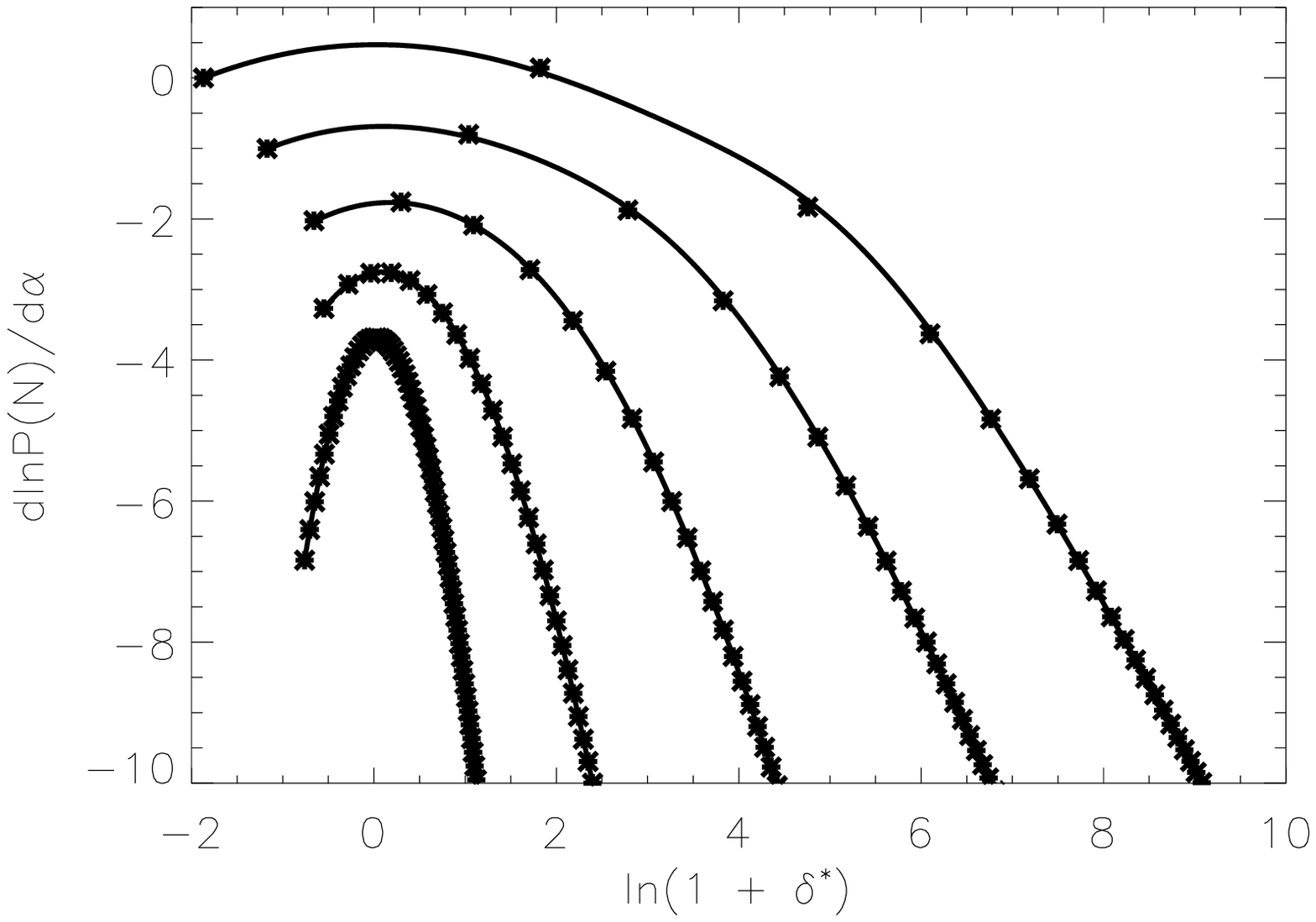}
\caption{Upper panel: the crosses show the exact count-in-cells PDF $P(N)$, as a function of $\ln(1 + \delta^*(N))$ for increasing values of the matter fluctuation variance as indicated from bottom to top. The corresponding number densities are given by the scaling \eqref{scaling}, with $\bar N_8 = 1, \sigma_8 = 0.8$.  The solid lines correspond to the saddle-point approximation to the PDF, Eq. \eqref{SP}. On each curve the first cross on the left indicate $P(0)$, the second $P(1)$, etc.  Lower panel: the same for the function $\partial_\alpha \ln P(N)$, giving the optimal observables. For our purposes, the saddle-point approximation is essentially exact.}
\label{figDev}
\end{center}
\end{figure}
The exact evaluation of $P(N)$ with numerical methods poses no particular difficulties. We used a basic Newton-Raphson algorithm to solve Eq. \eqref{saddlepoint} for the saddle point. We show $P(N)$ versus $A^*(N)$ as the crosses on the lower panel of Figure~\ref{figDev}, for $\bar N_8 = 1$, for different values of $\sigma$ as indicated. The solid lines show the saddle-point approximation. For convenience these lines omit the discreteness of the PDF, we continued $N$ to non-integer values by replacing the factorial function $N!$ by the Gamma function $\Gamma(N+1)$.  We found the relative deviation to be always subpercent even for large values of $\sigma$, except at the void probability $P(N = 0)$, where the accuracy worsen as $\sigma$ increases. The accuracy is similar or better for $\bar N_8 = 0.2$. It turns out that for all the purposes of this paper, the saddle-point approximation is as good as the exact result.
The lower panel shows the exact sufficient observable $\partial_\alpha \ln P(N)$ and its approximation with $o(N)$, Eq. \eqref{Om}. It is clear from the upper panel  that the transformation from $N$ to $A^*(N)$ no longer Gaussianizes the PDF on small scales. Nevertheless, the shape of $\partial_\alpha \ln P$ remains close to a parabola on all scales. This suggests that the entire information is 
contained the first two moments, regardless of the non-Gaussianity of $P(A^*)$. 
This observation will be fully exploited next.
\subsubsection*{Efficiencies}
 \begin{figure}
\begin{center}
\includegraphics[width = 0.5\textwidth]{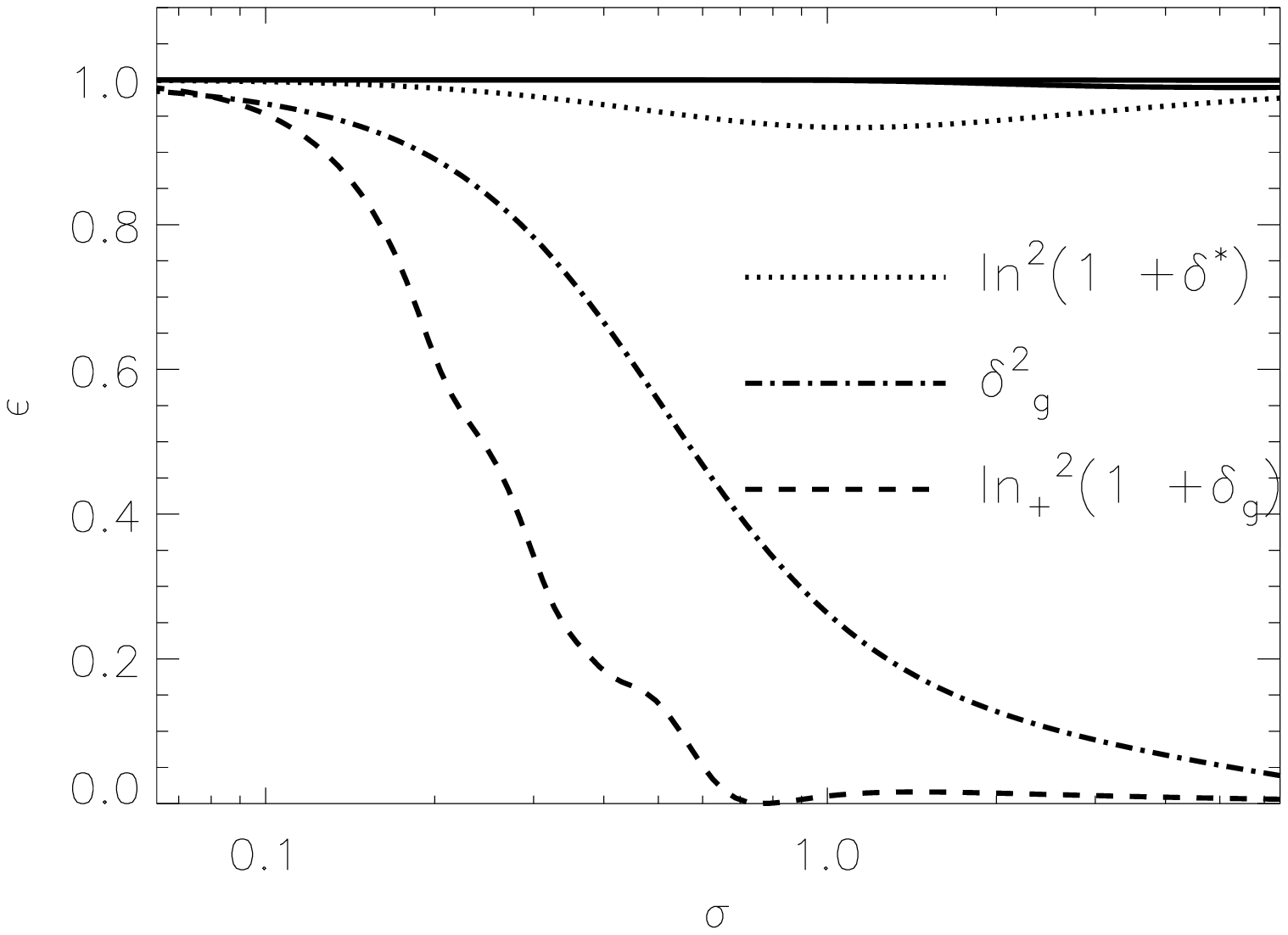}
\includegraphics[width = 0.5\textwidth]{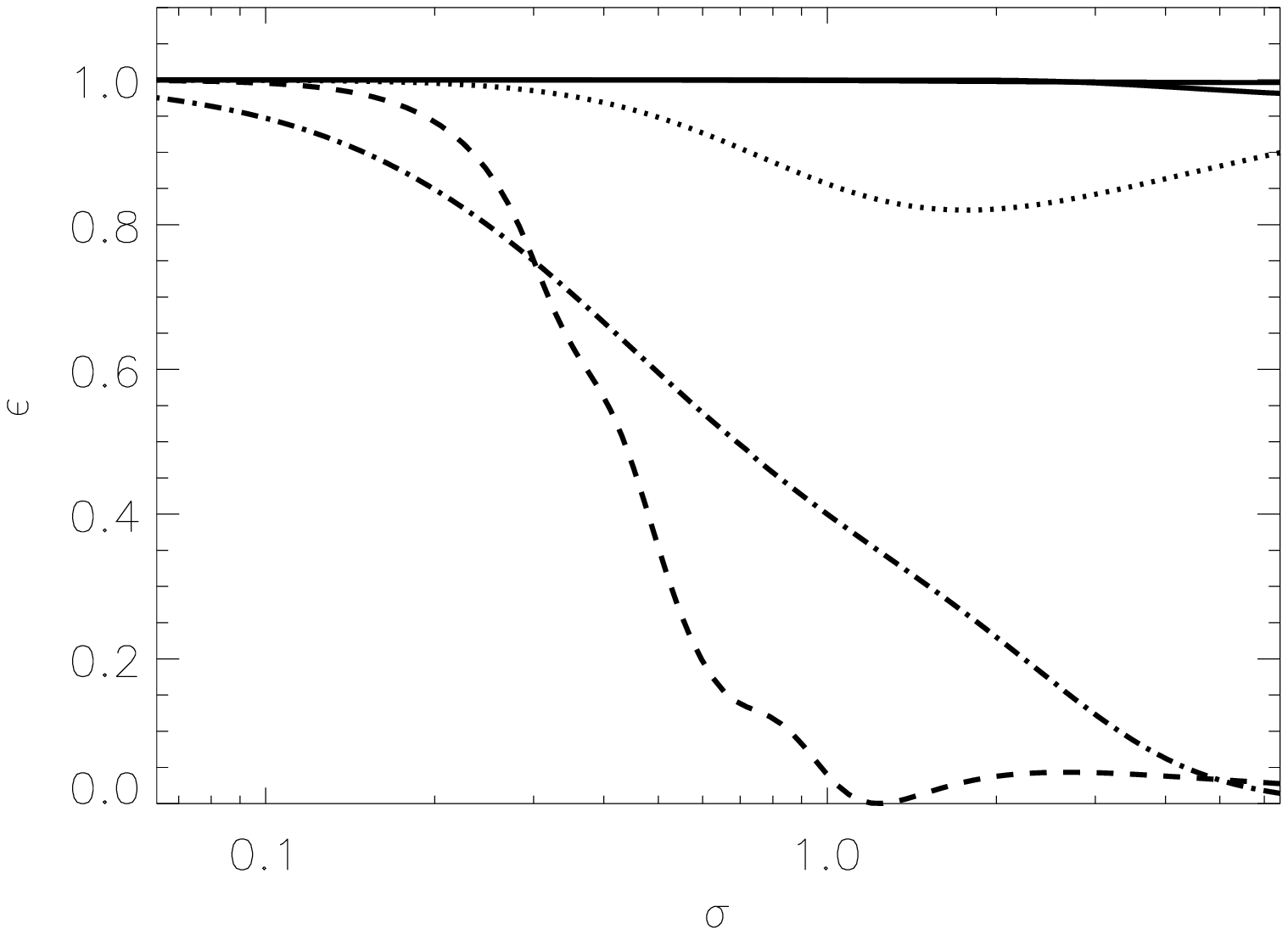}
\caption{Upper panel : the efficiency of various statistics to capture the information within the count-in-cell PDF, as a function of $\sigma$. $\bar N$ scales according to Eq. \eqref{scaling}, with $\bar N_8 = 0.2$.  The dotted line shows $\ln^2(1 + \delta^*)$, the dashed line $\log^2_+(1 + \delta_g)$, as defined in the text. The dash-dotted line shows the second moment of the galaxy density fluctuations $\delta_g^2$. Two other statistics are shown as solid lines, both almost indistinguishable from unity: the optimal statistics $o(N)$ according to the saddle-point approximation, and the combined information content of the first two moments of $\ln ( 1+ \delta^*)$, accounting for their covariance. The former shows a very slight deviation from optimality for the largest variances. The lower panel is the same for $\bar N_8 = 1$. In both cases, the optimal statistic is equivalent to the extraction of the first two moments of $\ln ( 1+ \delta^*)$.}
\label{figEff}
\end{center}
\end{figure}
 We define the efficiency of the statistics as the ratio $\epsilon$ of the information of the statistics to the total information. This is shown in figure \ref{figEff}, for two different  values of $\bar N_8$. Each panel shows several curves. 
 The lower solid line is the optimal statistic $o(N)$ according to our calculations, Eq. \eqref{Om}.
For the most part, it is indistinguishable from unity, confirming our expectations. A slight deviation is observed on the smallest scales, due to the the saddle-point approximation. The dotted curves show the efficiency of the second moment $\ln^2\lp 1 + \delta^*\rp$ alone, neglecting the curvature term in the optimal statistic; the latter still accounts for up to 20\% of the total information in the intermediate regime. The dash-dotted lines displays the efficiency of the naive $\delta_g^2$.
 This performs poorly for large variances, as expected. For comparison, we also show with dashes the statistic
 \beq
 \log^2_{+}(\delta) = \begin{cases} \ln^2(1 + \delta_g) & \delta_g > 0 \\  \delta_g^2 & \delta_g \leq 0 \end{cases} 
 \enq
 where the $\log_+$ mapping was introduced in \cite{NeyrinckEtal11} as local transform alternative to the logarithmic mapping that is well defined when $\delta_g = -1$. While useful on large scales when $\bar N$ is still sufficiently large, it performs poorly on smaller scales. In fact, its sensitivity to $\alpha$ vanishes around $\sigma \simeq 1$.
 \newline
 \indent
 Finally, motivated by the parabolic shape of $\partial_\alpha \ln P$ on the lower panel of figure \ref{figDev}, the upper solid lines present the efficiency of the first two moments (jointly) of $\ln(1 + \delta^*)$. Strikingly, these curves are indistinguishable from unity, showing perfect extraction of information. This suggests that the curvature term in $o(N)$ gleans the information content (perhaps in a complex way) of the first moment that is negligible for small, but carries information for larger $\sigma$. Note that this simple procedure appears to perform slightly better that $o(N)$ for the largest variances only due to the slight inaccuracy of the saddle point integration used to realise the optimal sufficient statistic. 

\section{Discussion} \label{discussion}
We have shown that to first order, the optimal statistics of a discrete field are given by that of an underlying field at the corresponding point $\delta^*$. Since the latter statistic is close to $\ln^2(1 + \delta)$, this suggests that the spectrum of $\ln (1 + \delta^*)$ is the key observable of the galaxy density field. Furthermore, our results suggest that the additional extraction of the mean captures the entire information. Our results for sufficient statistics are thus recast in terms of an optimal non-linear transformation. As we show next, $\delta^*$ has a meaningful physical interpretation. For the derivation of the optimal statistic, $\delta^*(N)$, or $A^*= \ln (1 + \delta^*)$, was introduced as a convenient mathematical construct: the sufficient statistic of the matter field needs to be evaluated at this point. We now reinterpret these results in
terms of reconstructing the underlying $A$ field from the observation of $N$. In a Bayesian setting, the posterior for $A$ is
\beq
p(A | N) = \frac{P(N | A) p(A)}{P(N)}.  
\enq  
By definition $A^*(N)$ maximizes  $\ln P(N|A) + \ln p(A)$, thus the right hand side of the above equation with respect to $A$. Therefore we can now interpret $A^*$ is the maximum a posteriori (MAP) solution in a Bayesian reconstruction of the $A$ field. The generalization of these one-dimensional considerations to the multipoint case using the full joint probability $P(N_1,\cdots, N_{\textrm{Ncell}})$  is clear. Statistics optimal with respect to $\alpha$ are given now by
\beq
\frac{\partial \ln p(A^*_1,\cdots,A^*_{\textrm{Ncell}})}{\partial \alpha} - \frac 12 \frac{\partial \ln \det H}{\partial \alpha},
\enq
where $H_{ij}$ is the curvature matrix  $-\partial^2_{ij}  \ln p(A_1,...| N_1,...)  $ evaluated at $A^*$,
 the saddle-point solution for the $A$ field. These coupled equations result in general in a more complicated, non-local transformation of the galaxy field.
The Bayesian reconstruction of the matter field with a Poisson lognormal assumption is the exact approach used in \cite{KitauraEtal10}. According to the above, their Bayesian MAP equations are always identical to our saddle-point equations. They implemented and solved these equations in three dimensions, which proves the feasibility of our approach even in the more involved multipoint case. On the other hand, their good overall success in reconstructing the density field and the direct connection we revealed between the MAP solution and sufficient statistics suggest that the logarithmic transformation of this reconstructed field does capture most information in the data. 
\newline
\indent
Our initial theoretical results suggest recipes for efficient observables from galaxy fields. The simplest is the following: map in each cell $N$ to $A^*(N)$ and extract the spectrum and mean of the new field. The mapping requires only an estimate of $\sigma^2_A$ and $\bar N$ that can be obtained from count-in-cells. There are some obvious generalizations and refinements that are left for future work. The multipoint case, including the curvature term, can be implemented with straightforward extensions of the
algorithms of \cite{KitauraEtal10}.  Further, these methods should be tested in simulations to see to what extent the 
conclusions from the lognormal model hold in real dark matter distributions. Refinements of the statistical models can include the use of the exact sufficient observable of the matter field and/or using a more accurate PDFs. Deviations from Poisson sampling and diverse biasing schemes can be modeled as well.

\section*{Acknowledgments}
We thank Mark C. Neyrinck for his careful reading of the manuscript, triggering improvements to the paper.
We acknowledge NASA grants NNX12AF83G and NNX10AD53G for support.

\bibliographystyle{mn2e}
\bibliography{bib}

\end{document}